\documentclass[preprint,showpacs,preprintnumbers,amsmath,amssymb]{revtex4}
\usepackage{graphicx}
\usepackage{dcolumn}
\usepackage{bm}
\usepackage{amsfonts}%
\usepackage{amsmath}

\begin{document}


\title{Calculation of three-body resonances using slow-variable discretization coupled with complex absorbing potential}
\author{Juan Blandon$^{1}$, Viatcheslav Kokoouline$^{1}$, Fran\c coise Masnou-Seeuws$^2$}
\affiliation{$^1$ Department of Physics, University of Central Florida, Orlando, Florida 32816, USA \\ 
$^2$ Laboratoire Aim\'e Cotton, CNRS, B\^at 505, Campus d'Orsay,  91405 Orsay Cedex, France}

\date{\today}

\begin{abstract}

We developed a method to calculate positions and widths of three-body resonances. The method combines the hyperspherical adiabatic approach,  slow variable discretization method (Tolstikhin {\it et al.}, J. Phys. B: At. Mol. Opt. Phys.  {\bf 29}, L389 (1996)), and a complex absorbing potential. The method can be used to obtain resonances having short-range or long-range wave functions. In particular, we have applied the method to obtain very shallow three-body Efimov resonances for a model system (Nielsen {\it et al.}, Phys. Rev. A {\bf 66}, 012705  (2002)).
\end{abstract}

\pacs{31.15.Ja, 02.70.-c, 33.20.Tp}

\maketitle

\section{Introduction}

The problem of weakly bound states of three resonantly interacting particles, first discussed long ago by Efimov \cite{efimov1971}, has attracted a particular interest in recent years \cite{esry96,braaten2002,stoll2005,esry2006,kraemer2006} owing to the development of refined experimental techniques allowing tunning of the inter-particle interaction. Now, it has become possible to tune the two-body interaction in such a way that a  two-body  weakly bound or virtual state appears. In this situation, three particles interacting pairwise through the two-body potential have a series of weakly bound states \cite{efimov1971}. If it happens that in addition to the weakly bound or virtual two-body state, there is another two-body bound state  lying deeply below,  the Efimov states are able to dissociate into a dimer and a free atom. Therefore, in such a situation, the Efimov states have finite lifetimes : these are Efimov resonances \cite{esry2006,kraemer2006,penkov99,nielsen2002}. Recently, such Efimov resonances have been observed in experiment \cite{kraemer2006}. 

Efimov predicted an universal behavior for the energies of the Efimov states \cite{efimov1971}: If one knows the position, let's say, of the lowest state, then the others can be found using a scaling law
\begin{equation}
E_{n}=E_{n-1}\exp(2\pi/\xi)\,,
\end{equation}
where $\xi$ is a constant depending on the actual interaction potential. This is possible because the long-range potential of an Efimov-type system behaves as $(\xi^2+0.25)/(2\mu\rho^2)$, where $\rho$ is the hyper-radius and $\mu$ is the reduced three-body mass. The long-range behavior of the potential does not depend on details of the short-range two-body interaction. For the same reason, widths of the Efimov resonances are scaled in the same way \cite{penkov99,nielsen2002}. Thus, if one needs to find positions and widths of the Efimov resonances for a particular three-body system, one has to be able to obtain at least one state: The approximate positions (and widths) of others can be obtained using the scaling law. 

Wave functions for such weakly bound Efimov states are  characterized by a probability density located mainly at large inter-particle distances. On the other hand, in order to correctly obtain the accumulated phase, the small-distance part of the wave functions must also be represented accurately, even though  the amplitude at  such distances can be very small. Due to the very different lengthscales in the two regions,  the calculation of the wave functions (of bound or resonant states) becomes difficult. 

Pen'kov \cite{penkov99} previously presented a theory of Efimov resonances using the formalism of Faddeev equations. He calculated six resonances for a model two-body interaction potential: the obtained positions and widths are in agreement with the scaling law. Later,  Nielsen {\it et al.} \cite{nielsen2002}  computed Efimov resonances for a different model potential, using an adiabatic hyperspherical method  coupled with a multi-channel $R$-matrix method along the hyper-radius.   In this approach, the authors obtained the scattering matrix as a function of energy above the lowest dissociation threshold. Then,  they deduced widths and positions of the resonances from the variation of the phase shift $\delta(E)$ in the scattering matrix.

Here, we are suggesting a different approach to calculate widths and positions of three-body resonances. This approach is general and can be used for weakly bound (long-range) and deeply-bound (short-range) resonances. The method   uses hyperspherical coordinates similar to Ref. \cite{nielsen2002}. Instead of solving a multi-channel scattering problem, we suggest employing the slow variable discretization (SVD)  \cite{tolstikhin96} method with adapted grid step \cite{kokoouline2006} in hyper-radius and hyperangles. As a basis set in hyper-radius we use the sine-DVR (discrete variable representation) method \cite{borisov01,willner2004}. At a large hyper-radius, we place a complex absorbing potential (CAP) to absorb the dissociating flux. Positions and widths of the three-body resonances are obtained  as real and imaginary parts of complex eigenvalues of the resulting Hamiltonian of the system.

The article is organized in the following way. In section \ref{sec:hyper} we present our approach of calculation of the three-body resonances. Then, in section \ref{sec:barier}, we apply the method to obtain a single resonance of three particles interacting through a potential with a barrier \cite{fedorov03}. The wave function of the resonance has a large amplitude mainly at relatively short distances. The second application of the developed method is discussed in section \ref{sec:Efimov}, where we calculate several Efimov resonances for three particles interacting through a two-body short-range potential with a large two-body scattering length. We compare our results with a previous theoretical study \cite{nielsen2002}. Finally, in section \ref{sec:concl}, we discuss advantages and possible applications of the method.

\section{Method}
\label{sec:hyper}

\subsection{Hyperspherical coordinates used in the present method}
In our approach we assume that the total angular momentum of the three-body system is zero. Thus, only three interparticle coordinates determine the configuration of the system. We represent the three interparticle coordinates by the Smith-Whitten hyperspherical coordinates \cite{johnson80}. The formulas of the actual version of the coordinates can be found in Refs. \cite{kokoouline2006,johnson80}. We use the same notations as Ref. \cite{kokoouline2006}. For a graphical view of the coordinates  we refer the reader to Fig. 6 of Ref. \cite{kokoouline03} and Fig. 1 of Ref. \cite{kokoouline2006}. In those studies the hyperangle $\phi$ changes in the interval $[0;2\pi]$. For systems with three identical particles (symmetry group $C_{3v}$), which are only considered in this study, the possible irreducible representations are $A_1$, $A_2$, and $E$. In order to represent  $A_1$ or $A_2$ wave functions, it is enough to consider a smaller interval of $\phi$, namely $[\pi/6;\pi/2]$ (see Figs. 6  and 7 of Ref. \cite{kokoouline03}). Wave functions of the $E$ irreducible representation require a larger interval $[\pi/6;7\pi/6]$. The systems considered here have $A_1$ resonances only. However, the method works for other irreducible representations and can be adapted to systems with non-identical particles. The second hyperangle $\theta$ changes in the interval $[0;\pi/2]$. The shape of a three-particle configuration is determined by the two hyperangles, the overall size of the system is determined by the hyper-radius $\rho\in[0;\infty)$. Therefore, the hyper-radius is the natural dissociation coordinate for the three-body system.

\subsection{The adiabatic hyperspherical approximation}
\label{sub:ada_accuracy}

For a three-body system interacting through the potential $V(\rho,\theta,\phi)$, we have to solve the Schr\"odinger equation 
\begin{equation}
\label{eq:Schr1}
\left[\hat{\cal T}(\rho,\theta,\phi)+V(\rho,\theta,\phi)\right]\Phi_{n}(\rho,\theta,\phi)=E^{vib}_n\Phi_{n}(\rho,\theta,\phi)\,
\end{equation}
to obtain the vibrational three-body wave function $\Phi_{n}(\rho,\theta,\phi)$ and eigenenergy $E^{vib}_n$. In this study, $V(\rho,\theta,\phi)$ is given by a sum of two-body potentials. The present method does not use that property of the potential and works for potentials that are not separable into a simple sum of the three two-body terms. Usually, realistic three-body potentials cannot be represented in such a way. 

One way of solving Eq. (\ref{eq:Schr1}) is to use the hyperspherical adiabatic approach, treating the hyper-radius as an adiabatic coordinate. Then, Eq. (\ref{eq:Schr1}) is solved in a two-step procedure. In the first step, one 'freezes' the hyper-radius at $\rho_i$ and solves the Schr\"odinger equation in  a two-dimensional space of hyperangles 
\begin{equation}
\label{eq:Had}
H^{ad}_{\rho_i} \varphi_a(\rho_i;\theta,\phi)=U_a(\rho_i)\varphi_a(\rho_i;\theta,\phi),
\end{equation}
where $\varphi_a(\rho_i;\theta,\phi)$ is the hyperangular wave function at hyper-radius $\rho_i$ for the $a^{th}$ eigenvalue $U_a(\rho_i)$ (see, for example, Fig. \ref{fig:hpawfs0001}). The Hamiltonian $H^{ad}_{\rho_i} $ is obtained from the full Hamiltonian of Eq. (\ref{eq:Schr1}) by fixing the hyper-radius. The eigenenergies $U_a(\rho_i)$ are called adiabatic potentials. Once the $U_a(\rho_i)$ are obtained, Eq. (\ref{eq:Schr1}) is written in the form of a system of coupled differential equations (see, for example, Ref. \cite{davydov})
\begin{equation}
\label{eq:non_ad}
\left[\hat T(\rho)+U_a(\rho)\right]\psi_{a,n}(\rho)+\sum_{a'}\left[ W_{a,a'}(\rho)\psi_{a',n}(\rho)\right] =E^{vib}_{n}\psi_{a,n}(\rho),
\end{equation}
where $T(\rho)$ is the kinetic energy operator associated with hyper-radial motion, $W_{a,a'}(\rho)$ are the coupling terms between the different channels $a$ and $a'$, and $\psi_{a,n}(\rho)$ is the $a^{th}$-channel component of the hyper-radial wave function. Assuming that the complete set of adiabatic channels $a$ is included in Eq. (\ref{eq:non_ad}), the solutions of Eqs. (\ref{eq:Schr1}) and (\ref{eq:non_ad}) are equivalent. Sharp $\rho$-dependence of the coupling terms $W_{a,a'}(\rho)$ makes the numerical solution of Eq. (\ref{eq:non_ad}) difficult.

\subsection{Mapped DVR method and slow variable discretization}
\label{sec:SVD}

The slow variable discretization method suggested by Tolstikhin {\it et al.} \cite{tolstikhin96} is used to account for the non-adiabatic coupling terms between the adiabatic states $\varphi_a(\rho_i;\theta,\phi)$ in a different way. We give only the formulas that are used in the current work. The detailed discussion of SVD can be found in Ref. \cite{tolstikhin96}. The total vibrational eigenstate $\Phi(\rho,\theta,\phi)$ (index $n$ is omitted here) is represented as an expansion in the basis formed by the adiabatic states $\varphi_a(\rho_i;\theta,\phi)$ obtained from Eq. (\ref{eq:Had}). The expansion coefficients are $\psi_{a}(\rho_i)$:
\begin{equation}
\label{eq:vibr_func_SVD}
\Phi(\rho,\theta,\phi)=\sum_a  \psi_{a}(\rho_i) \varphi_a(\rho_i;\theta,\phi)\,,
\end{equation}
where the sum is over all channels $a$. Expanding the hyper-radial wave function  in a complete basis set $\pi_j(\rho)$
\begin{equation}
\psi_{a}(\rho)=\sum_j  c_{j,a}\pi_j(\rho)\,,
\end{equation}
we write the hyper-radial Schr\"odinger equation, Eq. (\ref{eq:non_ad}), in the form of
\begin{equation}
\label{eq:gener_eigen}
\sum_{i',a'}\Big[\langle\pi_{i} |\hat T(\rho)|\pi_{i'}\rangle{\cal O}_{ia,i'a'}+\langle\pi_{i}|U_a(\rho)|\pi_{i'}\rangle\delta_{aa'}\Big]c_{i'a'}=E\sum_{i',a'}\langle\pi_{i}|\pi_{i'}\rangle{\cal O}_{ia,i'a'}c_{i'a'}\,,
\end{equation}
where the  overlap matrix elements ${\cal O}_{ia,i'a'}$ are 
 \begin{equation}
{\cal O}_{ia,i'a'}=\langle\varphi_{a}(\rho_{i};\theta,\phi)|\varphi_{a'}(\rho_{i'};\theta,\phi)\rangle\,.
\end{equation}
These matrix elements account for the non-adiabatic couplings between the channels. They replace the $\rho$-dependent non-adiabatic terms $W_{a,a'}(\rho)$ in Eq. (\ref{eq:non_ad}). The above Eq. (\ref{eq:gener_eigen}) is written in a form of a generalized eigenvalue problem, which can be solved using standard numerical procedures. For an orthonormal DVR basis $\pi_i$, the equation is finally reduced to
\begin{equation}
\label{eq:gener_eigen2}
\sum_{i',a'}\left[\langle\pi_{i} |\hat T(\rho)|\pi_{i'}\rangle{\cal O}_{ia,i'a'}+U_a(\rho_i)\delta_{ii'}\delta_{aa'}\right]c_{i'a'}=E\sum_{a'}{\cal O}_{ia,ia'}c_{ia'}.
\end{equation}
In the above equations, the kinetic energy operator $\hat T(\rho)$ is
\begin{equation}
\hat T(\rho)=-\frac{\hbar^2}{2\mu}\frac{d^2}{d\rho^2}.
\end{equation}

We use a DVR basis $|\pi_i\rangle$ to represent the hyper-radial part of the wave functions. Since the weakly bound states and resonances have wave functions extending to large distances, we use a variable grid step for DVR basis functions similarly as we did in Refs. \cite{kokoouline2006,kokoouline99}. Namely, we introduce a new variable $x$ such that $\rho=\rho(x)$. The grid steps $\Delta\rho$ and $\Delta x$ along $\rho$ and $x$, correspondingly, are linked to each other as  $\Delta\rho=J(x)\Delta x$, where $J(x)=d\rho/dx$ is the Jacobian. We may set $\Delta x$ to be 1 and, therefore, $\Delta\rho=J(x)$. In practice, first, we choose the dependence $\Delta\rho(\rho)$. Once $\Delta\rho(\rho)$ is chosen, the grid step in $\rho$ automatically determines $J(x)$. Then the integral $\int J(x)dx$ gives the dependence $\rho(x)$.  Changing the variable $\rho$ to $x$ in the kinetic energy operator $\hat T(\rho)$ and the wave function $\psi$ to $\tilde\psi=\sqrt{J}\psi$, Eq. (\ref{eq:gener_eigen2}) becomes
\begin{equation}
\label{eq:gener_eigen3}
\sum_{i',a'}\tilde T_{ii'}{\cal O}_{ia,i'a'}\tilde c_{i'a'}+U_a(\rho(x_i))\tilde c_{ia}=E\sum_{a'}{\cal O}_{ia,ia'}\tilde c_{ia'}\,,
\end{equation}
where $\tilde c_{ia}=\sqrt{J(x_i)} c_{ia}$. The matrix of the kinetic energy operator $\tilde T(x)$ is determined once the DVR basis set $\pi_i(x)$ is chosen. Previously, we used complex exponents, $\exp({\rm i} k x_i)$ as DVR basis functions. However, as it was pointed out in Refs. \cite{borisov01,willner2004}, the sine function basis, $\sin({\rm i} k x_i)$ has a certain advantage: zero boundary conditions at the ends of the grid are better represented by the sine functions than by the complex exponents. So, in this study we use the sine DVR basis. The actual form of the orthonormal DVR sine basis functions $\pi_i(x)$ is well known. See, for example, Eqs. (5-7) of Ref.  \cite{borisov01}. To calculate the matrix elements  $\tilde T_{i'i}$ of the kinetic energy operator we use an approach similar to the one described in Refs.\cite{borisov01,willner2004}. The matrix $\tilde T_{ii'}$  can be represented as the following matrix product
\begin{eqnarray}
 {\tilde  T}=\frac{\hbar^2}{2\mu}\left[\hat B\hat A_{cos}^{-1}\hat K \hat A_{sin}\hat R \hat A_{sin}^{-1}\hat K\hat A_{cos}\hat B\right]\,,
\end{eqnarray}
where $B$, $R$, and $K$ are diagonal matrices \cite{borisov01}
\begin{eqnarray}
B_{ij}=\delta_{ij}/\sqrt{J(x_i)}\,, R_{ij}=\delta_{ij}/J(x_i)\,,\nonumber\\
K_{ij}=\delta_{ij}\frac{\pi i}{N}\,.
\end{eqnarray}
$N$ is the total number of the DVR basis functions. The elements of the matrices $\hat A_{sin}$, $\hat A_{sin}^{-1}$, $\hat A_{cos}$, $\hat A_{cos}^{-1}$ are \cite{borisov01,willner2004}
\begin{eqnarray}
\left(\hat A_{sin}\right)_{ij}=\sin\left(\frac{(2j-1)}{2N}i\pi\right)\alpha_i \,,\nonumber \\
\left(\hat A_{sin}^{-1}\right)_{ji}=\frac{2}{N}\sin\left(\frac{(2j-1)}{2N}i\pi\right)\alpha_i \,,\nonumber \\
\left(\hat A_{cos}\right)_{ij}=\cos\left(\frac{(2j-1)}{2N}i\pi\right)\alpha_i \,,\nonumber \\
\left(\hat A_{cos}^{-1}\right)_{ji}=\frac{2}{N}\cos\left(\frac{(2j-1)}{2N}i\pi\right)\alpha_i\,,
\end{eqnarray}
where
\begin{eqnarray}
\alpha_i=\left\{ \begin{array}{cc}
1/\sqrt{2}\,, & i=0,N\,,\\
1\,,& {\rm otherwise}\,,
\end{array}\right.\\
 {\rm and}\  i=0,1,\ldots N\,; j=1,\ldots N \,.\nonumber
\end{eqnarray}

To represent wave functions of loosely-bound states we choose a smaller step $\Delta\rho_i=J(x_i)$ in the regions where the wave functions oscillate a lot, and a larger grid step for the regions where the wave functions change smoothly. One possibility is to connect the grid step $\Delta\rho$ to the local kinetic energy $E_\mathrm{kin}(\rho)$ \cite{kokoouline2006,kokoouline99}: 
\begin{equation}
\label{eq:Delta}
\Delta\rho=\beta\frac{\pi\hbar}{\sqrt{2\mu E_\mathrm{kin}(\rho)}}\,.
\end{equation}
where the parameter $\beta \le 1$ provides additional flexibility. This  choice for  $\Delta\rho$ gives good results in the regions where the WKB applicability condition is satisfied. In other situations, around classical turning points, for example, one needs to take a smaller step.

\subsection{Complex absorbing potential}
\label{sec:CAP}

A complex absorbing potential is used to simulate an infinite hyper-radial grid (see Refs. \cite{vibok1992}, \cite{riss1993}  and references therein). The CAP is placed at the end of the grid and, therefore, absorbs the outgoing dissociation flux of decaying resonant states. The length and the strength of CAP are chosen to minimize its effect: the outgoing flux should be completely absorbed and should not be reflected back to the internal region. The CAP is purely imaginary and added to the adiabatic potentials $U_a(\rho)$. This makes the Hamiltonian non-Hermitian. Solving the generalized complex eigenvalue problem of Eq. (\ref{eq:gener_eigen}) the eigenenergies of the Hamiltonian are obtained in the form
\begin{equation}
\label{eq:eig_energies}
E = E' - i\frac{\Gamma}{2},
\end{equation}
where $E'$ is the position of the resonance and $\Gamma$ is the resonance linewidth (inverse of lifetime) \cite{riss1993}.

In literature, different types of CAPs are used  (see Refs. \cite{vibok1992}, \cite{riss1993} and references therein). In our calculations, we use  quadratic and exponential forms of CAP. The widths and positions of the resonances do not depend on the form of CAP if the length and strength are chosen appropriately (see Table \ref{table:CAP}). The optimal lengths and strengths of the CAPs are determined using the de Broglie wavelength of the dissociating products according to guidelines provided in Ref. \cite{vibok1992}. In this case, the dissociating products are a dimer plus a free atom. The kinetic energy of the dissociating products, from which we get the de Broglie wavelength, is the difference between the three-body bound state energy of the resonance, $E_{trimer}$, and the dissociation energy of the ground state adiabatic curve, $E_{diss}$ (see inset in Fig. \ref{fig:3Bpots}).

\subsection{Reduced grid and variable grid-step in 2D space of hyperangles}
\label{ssec:redgrid}

In this study, we are interested only in vibrational states of the $A_1$ irreducible representation. Such states can be represented by reduced interval $[\pi/6;\pi/2]$ of the hyperangle $\phi$ (examples of $A_1$ functions are given in Figs. \ref{fig:hpawfs0001} and \ref{fig:hpawfs001}). Using Neumann boundary conditions
\begin{equation}
\frac{\partial\varphi_a(\phi = \pi/6)}{\partial\phi} = \frac{\partial\varphi_a(\phi = \pi/2)}{\partial\phi} = 0\,,
\end{equation}
we are able to represent states of the $A_1$ irreducible representation. To represent states with $A_2$ symmetry, we would have to impose Dirichlet boundary conditions. Such states come up in systems of fermions, for example (assuming that only the vibrational part of the total wave function determines the symmetry of the state). To represent $E_a$ or $E_b$ states, we would need to change our $\phi$ interval to [$\pi$/6,$7\pi$/2] and our boundary conditions according to $E_a$ (symmetric, Neumann) or $E_b$ (antisymmetric, Dirichlet) representation.

\begin{figure}[h]
\includegraphics[width=15cm]{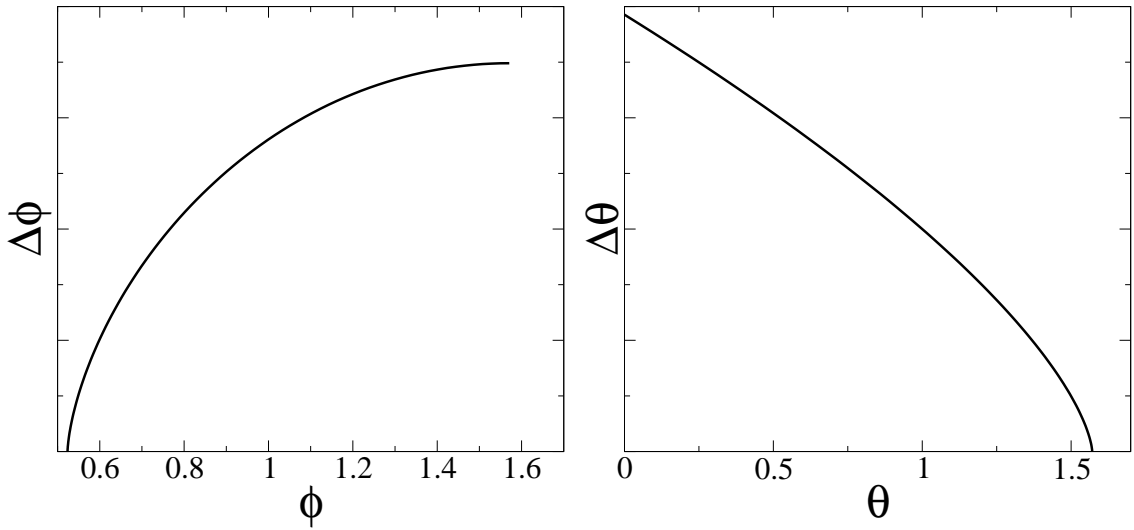}
\caption{Variable grid step-sizes, $\Delta\theta$ and $\Delta\phi$, for $\theta$ and $\phi$, respectively. For a more accurate representation of hyperangular wave functions (see Figs. \ref{fig:hpawfs0001}, \ref{fig:hpawfs001}), $\Delta\phi$ $\rightarrow 0$ as $\phi$ $\rightarrow$ $\pi/6$, whereas $\Delta$ $\theta$ $\rightarrow 0$ as $\theta$ $\rightarrow$ $\pi/2$.}
\label{fig:hpastep1}
\end{figure}

We also employ a variable grid in hyperangles because the hyperangular wave functions are extremely localized when the hyper-radius is large. This occurs particularly in the asymptotic region of hyper-radius, where the hyperangular wave function is nonzero only in a small hyperangular region representing the dimer plus free atom configuration. For example, in the upper-left frame of Fig. \ref{fig:hpawfs001}, we see the hyperangular wave function very localized near $\theta$ = $\pi$/2 and $\phi$ = $\pi$/6. The variable step size in the hyperangles, used in this study, is shown in Fig. \ref{fig:hpastep1}. The step size in $\theta$,  becomes smaller as $\theta$ approaches $\pi$/2, while the step size in $\phi$, approaches zero as $\phi$ approaches $\pi$/6. In principle, it is possible to use different dependences like those shown in Fig. \ref{fig:hpastep1} for different hyper-radii. It would save some computation time. However, in this study, we use the same hyperangular grid for all values of $\rho_i$.

\section{Example: three particles interacting through a two-body potential with a barrier}
\label{sec:barier}

\subsection{Our results}

To test our method, first, we applied it to a simple model system that consists of three bosons with nucleon mass, $m_1=m_2=m_3=1837.5773$ a.u. ($939$ MeV). The system was previously considered in Ref. \cite{fedorov03}, where the Faddeev equations were employed in order to obtain three-body states. The three-body potential $V(r_{12},r_{23},r_{31})$ in the Schr\"odinger equation, Eq. (\ref{eq:Schr1}), is constructed as a sum of pairwise potentials: $V(r_{12},r_{23},r_{31})=V_2(r_{12})+V_2(r_{23})+V_2(r_{31})$, where the pairwise interaction $V_2(r)$ is given by:
\begin{equation}
\label{eq:2body_barrier}
V_2(r)=-55 e^{-0.2 r^2}+1.5e^{-0.01(r-5)^2},
\end{equation}
in MeV and the distance $r$ is in fm. The two-body potential has a two-body bound state at $E^{(2)}$ = -6.76 MeV \cite{fedorov03}. 

\begin{figure}[ht]
\includegraphics[width=10cm]{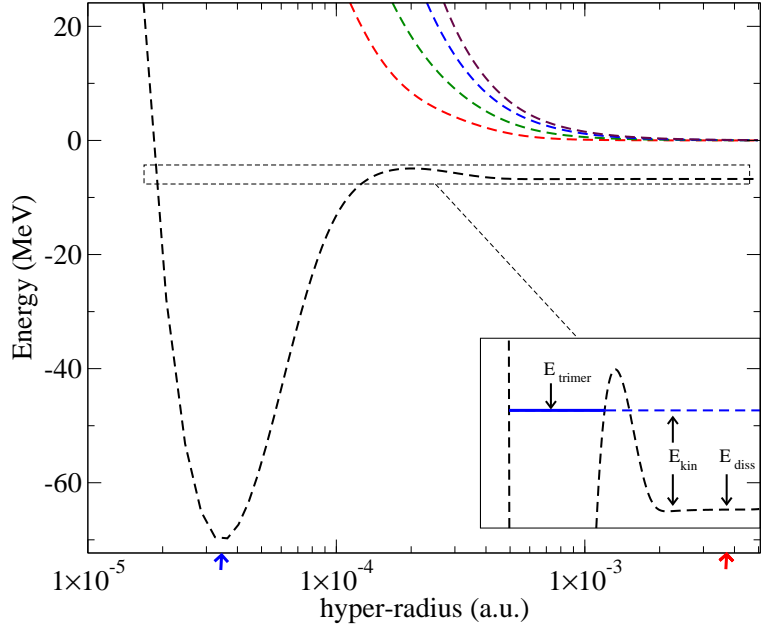}
\caption{(Color online) Hyperspherical adiabatic potentials for the system with a potential barrier. Five $A_1$ adiabatic channels are displayed. There is only one bound state at -37.35 MeV. Resonant states are long-living states that tunnel through the potential barrier as shown in the inset. Note logarithmic scale in hyper-radius. The two arrows along hyper-radius point to the values of $\rho$ where ground adiabatic curve has its minimum and to where CAP begins.}
\label{fig:3Bpots}
\end{figure}

As the first step, we obtain the adiabatic energy curves. The resulting adiabatic potentials are shown in Fig. \ref{fig:3Bpots}. The lowest potential at $\rho\to\infty$ corresponds to a decay into a bound pair of two nucleons and a free nucleon.  The dissociation energy of the ground state ($E_{diss}$ in Fig. \ref{fig:3Bpots}) corresponds to the dimer binding energy $E^{(2)}$.  All other channels dissociate into three free atoms. If the energy of a three-body state is above the lowest dissociation limit, the system may dissociate, with the excess of energy transferred into the kinetic energy of the dissociating products.

\begin{figure}[ht]
\includegraphics[width=10cm]{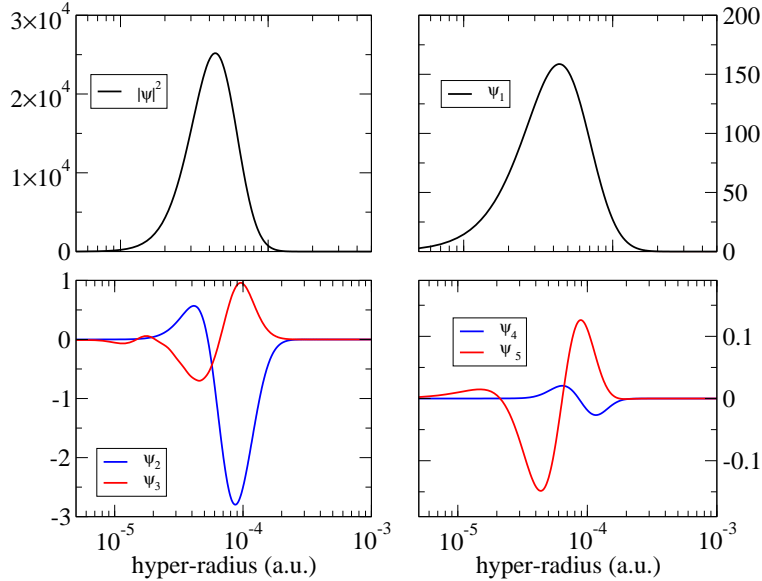}
\caption{(Color online) Ground state hyper-radial wave function, $\psi(\rho)$, broken down into components for five-channel calculation. Upper-left frame shows modulus squared of normalized wave function. Other three frames show channel by channel components, $a = 1$ is ground channel, $a = 2$ is first excited state channel, and so on. Note logarithmic scale in hyper-radius for all hyper-radial wave functions.}
\label{fig:gdstatewf}
\end{figure}

With our method, we are able to calculate three-body resonances and bound states. We find that there is only one three-body bound state at $E_0$ = -37.35 MeV. This result is in agreement with the previous study by Fedorov {\it et al.} \cite{fedorov03} giving $E_0$ = -37.22 MeV. Figure \ref{fig:gdstatewf} shows the multi-channel hyper-radial wave function $\psi(\rho)$ of this state.

The pairwise interaction potential has a barrier. It means that if we have three particles situated close to each other, the dissociation to the dimer + free particle configuration is separated by a potential barrier too. Thus the lowest adiabatic state has a potenital barrier too. If the barrier is high enough, one or more resonances (pre-dissociated states) may exist with the energy above the lowest dissociation limit (dimer + free particle). The energy interval between the dissociation limit and the top of the barrier is roughly from -7 MeV to -5 MeV.

\begin{figure}[ht] 
\includegraphics[width=10cm]{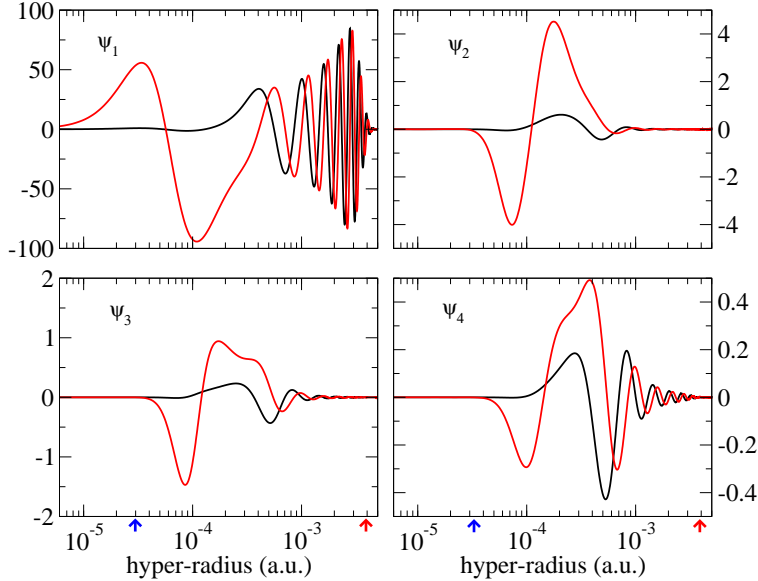}
\caption{(Color online) Resonance hyper-radial wave function for four-channel calculation. The wave function has a considerable amplitude in the region of the potential well (around $10^{-4}$ a.u.). Arrows along hyper-radius point to ground adiabatic curve minimum and to where CAP begins (see Fig. \ref{fig:3Bpots}). CAP is placed around $\rho=4\times10^{-3}$ a.u., the distance at which the outgoing dissociation flux starts to be absorbed. This can be compared with the continuum wave function shown in Fig. \ref{fig:contwf}, for which the potential-well amplitude is very small. Comparing the amplitudes of the different components, it is clear that the principal contribution is due to the lowest adiabatic state.}
\label{fig:reswf}
\end{figure} 

\begin{table}[tbp]
\vspace{0.3cm} 
\begin{tabular}{|p{4.5 cm}||p{4.0cm}|p{4.5cm}||p{3.0cm}|}
\hline
CAP strength (a.u.) & CAP length (a.u.)  & CAP initial position (a.u.)  & $E' - i\frac{\Gamma}{2}$ (MeV) \\ 
\hline
80,000                  &  0.0028             &  0.0060         &       -5.307 - i0.116  \\
\hline
104,000                 &  0.0028              &  0.0060        &     -5.305 - i0.114   \\
\hline
56,000                 &  0.0028             &  0.0060           &     -5.310 - i0.117  \\
\hline
80,000                  &  0.0036             &  0.0052         &       -5.318 - i0.115 \\
\hline
80,000                 &  0.0020              &  0.0068         &      -5.344 - i0.107  \\
\hline
\end{tabular}
\vspace{0.3cm}
\caption{Stability of resonance with respect to CAP parameters for system of identical bosons interacting through a two-body potential with a barrier. The deBroglie wavelength of the dissociating products is approximately 0.0006 a.u. The most deviant calculation, last row, is due to the short length of the CAP for that calculation.}
\label{table:stability}
\end{table}

\begin{figure}[ht]
\includegraphics[width=10cm]{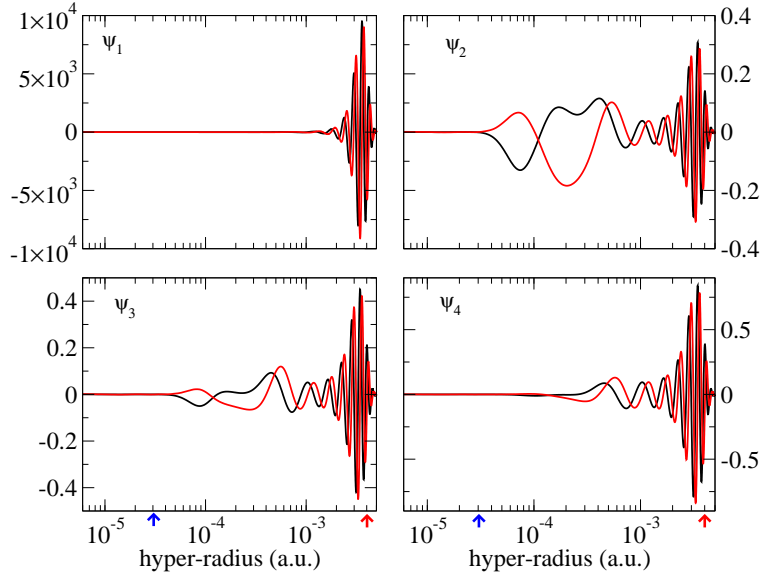}
\caption{(Color online) Channel by channel components of continuum state wave function for four-channel calculation ($E = -6.4$ MeV.). Compared to the resonant wave function, Fig. \ref{fig:reswf}, the amplitude of this wave function inside the potential-well region is negligible. Arrows along hyper-radius point to ground adiabatic curve minimum and to where CAP begins (see Fig. \ref{fig:3Bpots}).}
\label{fig:contwf}
\end{figure}

\begin{figure}[ht]
\includegraphics[width=10cm]{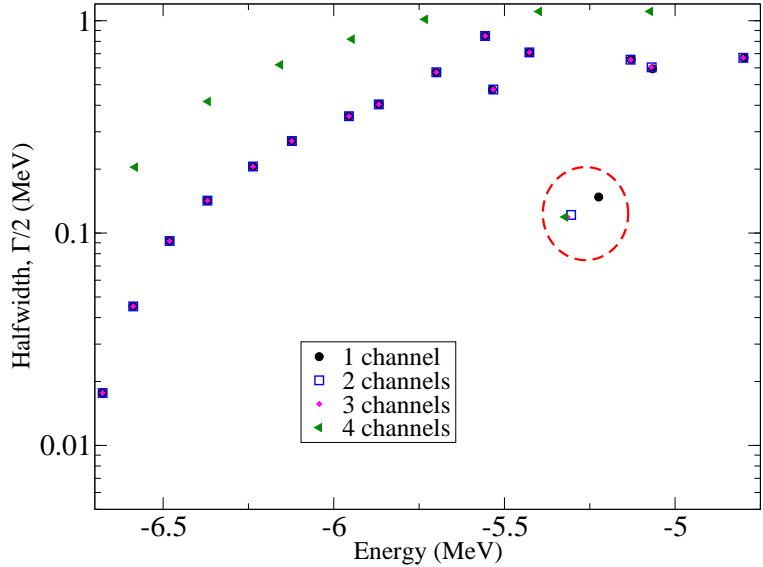}
\caption{(Color online) Convergence of the position and width of the resonance with respect to the number of adiabatic states included in the calculation. The results for the resonance are circled; other data points represent the continuum states. Positions and widths of the continuum states depend on the CAP parameters. Notice that positions of the resonance and continuum states are at negative energies because the origin of energy is at the three-body dissociation, which is above the dissociation to the dimer + free particle configuration.}
\label{fig:widthsvsE}
\end{figure}

Most of the eigenvalues of Eq. (\ref{eq:gener_eigen3}) are neither bound states nor resonances: these are just states of the continuum spectrum. To distinguish the resonances from the continuum states, we may look into their wave functions or carry out calculations for a different grid length and CAP parameters of the grid: Positions and widths of resonances should be stable with respect to the change of the grid length and CAP parameters. In our calculations, we used CAP length and strength equal to 0.0028 a.u. and 80,000 a.u., respectively. We found only one resonance for this system with energy -5.31 MeV and halfwidth 0.12 MeV. The stability of this resonance with respect to varying CAP parameters is illustrated in Table \ref{table:stability}. We varied CAP length and strength parameters by $+/- 30\%$ of optimal values. The wave function of the resonance is shown in Fig.  \ref{fig:reswf}. The ground state component of the hyper-radial wave function has non-negligible components both inside the bound region as well as in the continuum region, as expected for a resonant state. Note that the ground state component of the hyper-radial wave function, $\psi_1$, is at least one order of magnitude greater than any other component, meaning that this state 'lives' mostly in the ground adiabatic curve. Nevertheless, the presence of other components means that non-adiabatic couplings are not negligible. The significance of the non-adiabatic coupling is also demonstrated in Fig. \ref{fig:widthsvsE}, which shows the convergence of the results with respect to a different number of adiabatic channels included in Eq. (\ref{eq:gener_eigen3}): One channel is not enough to obtain an accurate result, two channels provide a much better accuracy. To compare with Fig. \ref{fig:reswf}, a continuum wave function for vibrational state with energy $E = -6.4$ MeV is shown in Fig. \ref{fig:contwf}. An important difference with the resonant wave function is that the continuum wave function is situated mainly in the outer region of the hyper-radius, not reaching the potential well.

For two-body systems, the complex energy of resonances does not depend on the shape of CAP, if CAP is smooth enough. So, we have also checked the convergence of our results with respect to different types of CAPs. Namely, we compared the quadratic and exponential CAPs. As expected, we found that the energy of the only resonance depends only weakly on the CAP parameters. Table \ref{table:CAP} compares the results obtained with the quadratic and exponential CAP.
\begin{table}[tbp]
\vspace{0.3cm} 
\begin{tabular}{|p{2.5 cm}||p{5.5cm}|p{5.5cm}|}
\hline
CAP type & Energy (MeV)  & Halfwidth (MeV) \\ 
\hline
quadratic     &  -5.319  &  0.1115 \\ 
\hline
 exponential  & -5.318   &  0.1183 \\ 
\hline
\end{tabular}
\vspace{0.3cm}
\caption{Stability of the calculated resonance with respect to type of CAP used.}
\label{table:CAP}
\end{table}

\subsection{Comparison with previous study}

Having developed the described method, we tested it on the present system because it was previously considered in Ref. \cite{fedorov03}. We obtained the energy of the only bound state, which is in agreement with the mentioned study. However, the position (-5.319 MeV)  and halfwidth (0.112 MeV) of the resonance do not agree with Ref.  \cite{fedorov03}, which gives position -5.96 MeV and halfwidth 0.40 MeV. This came as a surprise because the calculation for the second three-body system considered here is in a good agreement with a previous work \cite{nielsen2002}. The reason of the disagreement is not clear. We are confident about the adiabatic states obtained in this study because we have obtained the same bound state as Fedorov {\it et al.} We are confident about the one-channel calculation giving a value of the complex energy, which is quite close to the converged value. The one-channel code does not use SVD at all and we tested it on diatomic molecules with a barrier. 

\begin{figure}[ht]
\includegraphics[width=10cm]{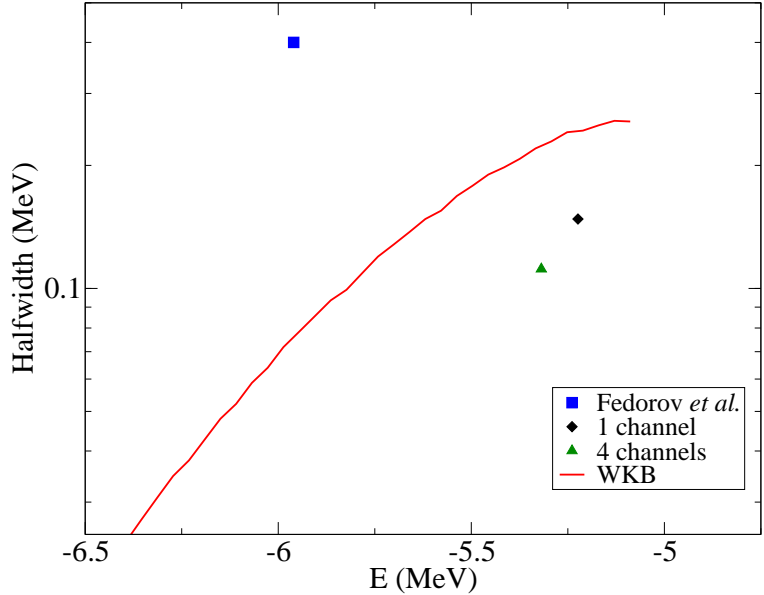}
\caption{(Color online) Comparison of our resonance linewidth calculations to linewidth calculation of Fedorov {\it et al.}, and to the WKB approximation. We show our halfwidths for one-channel and five-channel calculations. Our four-channel calculation is closer to WKB approximation than that of Fedorov {\it et al.} \cite{fedorov03}. However, WKB estimation cannot be viewed as reliable for this particular situation.}
\label{fig:widthscomparison}
\end{figure}

As an additional test, we carried out a calculation based on the WKB approximation. Strictly speaking, the WKB approximation can hardly be applied for this case, because the resonance is only the second state. WKB works only for excited states. But for the completeness of the discussion, we provide the details of the WKB calculation. In the WKB approximation, the linewidth is given by
\begin{equation}
\Gamma = \frac{P_T}{T}.
\end{equation}
The transmission probability $P_T$, and period $T$ of oscillations are given by \cite{tiemann86}
\begin{equation}
\label{eq:WKB1}
P_T = \ln\Big[1+\exp\left(-\frac{2}{\hbar}\int_b^c{\sqrt{2\mu \big\{U(\rho) - E_{trimer} \big\} } d\rho}\right)\Big],
\end{equation}
\begin{equation}
\label{eq:WKB2}
T = \int_a^b{\sqrt{\frac{2\mu}{E_{trimer} - U(\rho)} } d\rho},
\end{equation}
where $\mu$ is the three-body reduced mass, $E_{trimer}$ is the energy of the resonance shown in Fig. (\ref{fig:3Bpots}). For $T$, the limits of integration $a$ and $b$ are the classical turning points in the potential well. The limits of integration  $b$ and $c$ for $P_T$ are the 'entrance' and 'exit' points through which the three-body dissociating flux is tunneling. In Fig. \ref{fig:widthscomparison}, we compare our resonance linewidth to results previously obtained by Fedorov {\it et al.} \cite{fedorov03} and WKB estimation. Our results are somewhat closer to the WKB approximation than those from Ref. \cite{fedorov03}, although as pointed out above WKB estimation does not provide a reliable result for this situation.

One reason for the  discrepancy with Ref. \cite{fedorov03} could be that in our calculation we used a longer grid extending until $\rho_{max} \approx$ 0.01 a.u. than the grid in Ref. \cite{fedorov03}, where $\rho_{max} \approx$0.002 a.u. We found that to obtain converged results within our approach, $\rho_{max}$ should be at least $\approx$ 0.005 a.u. or larger: CAP should have a length long enough to absorb the dissociation flux smoothly (see Figs. \ref{fig:gdstatewf}, \ref{fig:reswf}, and \ref{fig:contwf}). In Ref. \cite{fedorov03}, the situation could be different because no CAP was used. Instead, a complex scaling variable was employed. However, the complex scaling variable plays a role of an absorber. If the length of the complex scaling variable is too short, the outgoing dissociation flux is not absorbed completely and, thus, is reflected back from the end of the grid. This should give a wrong value for the resonance width.

\begin{figure}[ht]
\includegraphics[width=10cm]{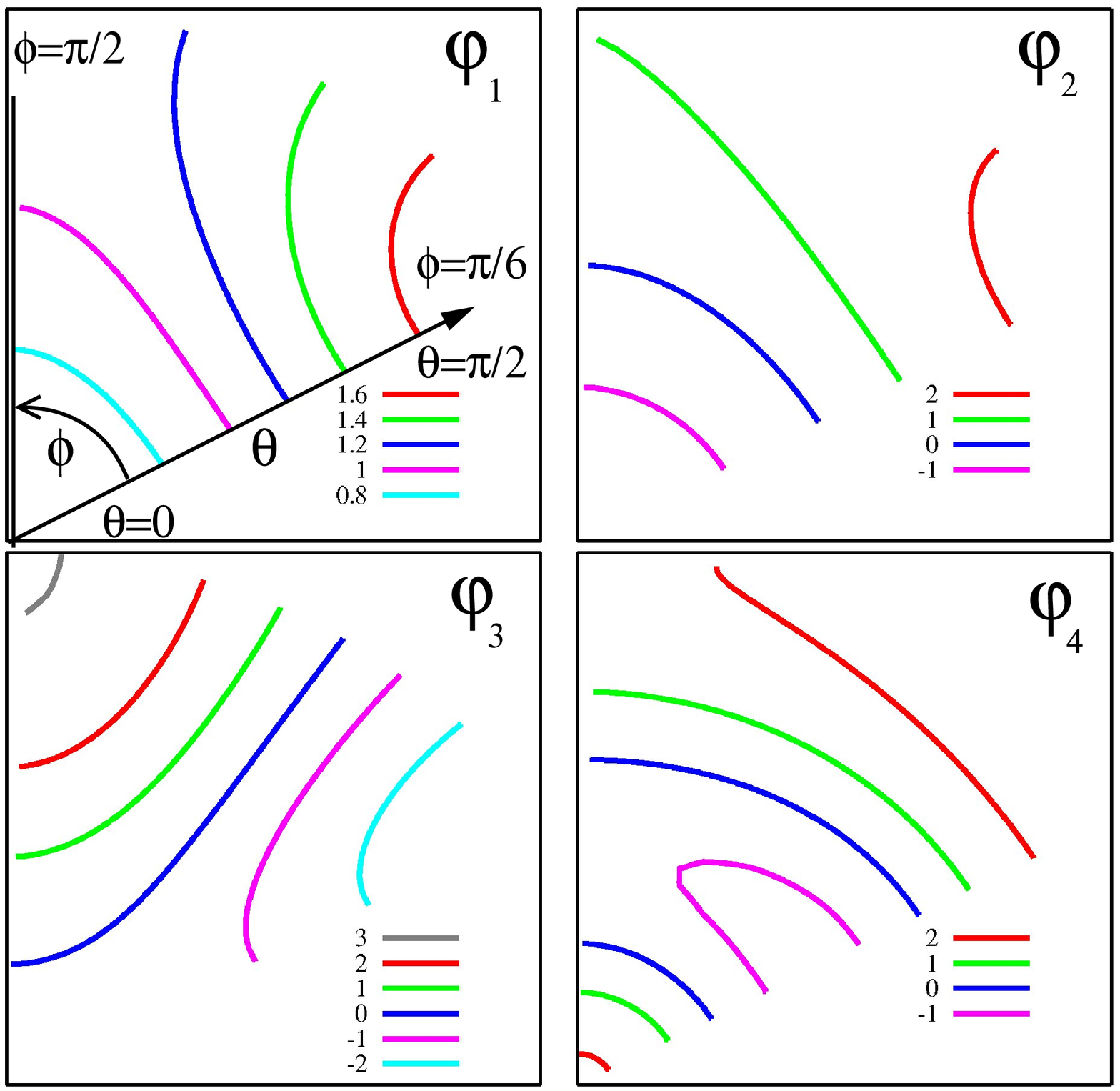}
\caption{(Color online) Contour plots of hyperangular wave functions at $\rho$ = 0.0001 a.u. for the system with potential given by Eq. (\ref{eq:2body_barrier}). Values of contour lines are given by the legends. Upper left frame shows hyperangular axes: hyperangle $\theta$ runs in the radial direction from $0$ to $\pi/2$, while $\phi$ runs in the polar angular direction from $0$ to $2\pi$. For this calculation we restict $\phi$ to $\pi/6 \le \phi \le \pi/2$. This value of $\rho$ corresponds to the bound state region. All wave functions are delocalized: there is no preferred three-body arrangement.}
\label{fig:hpawfs0001}
\end{figure}

\begin{figure}[ht]
\includegraphics[width=10cm]{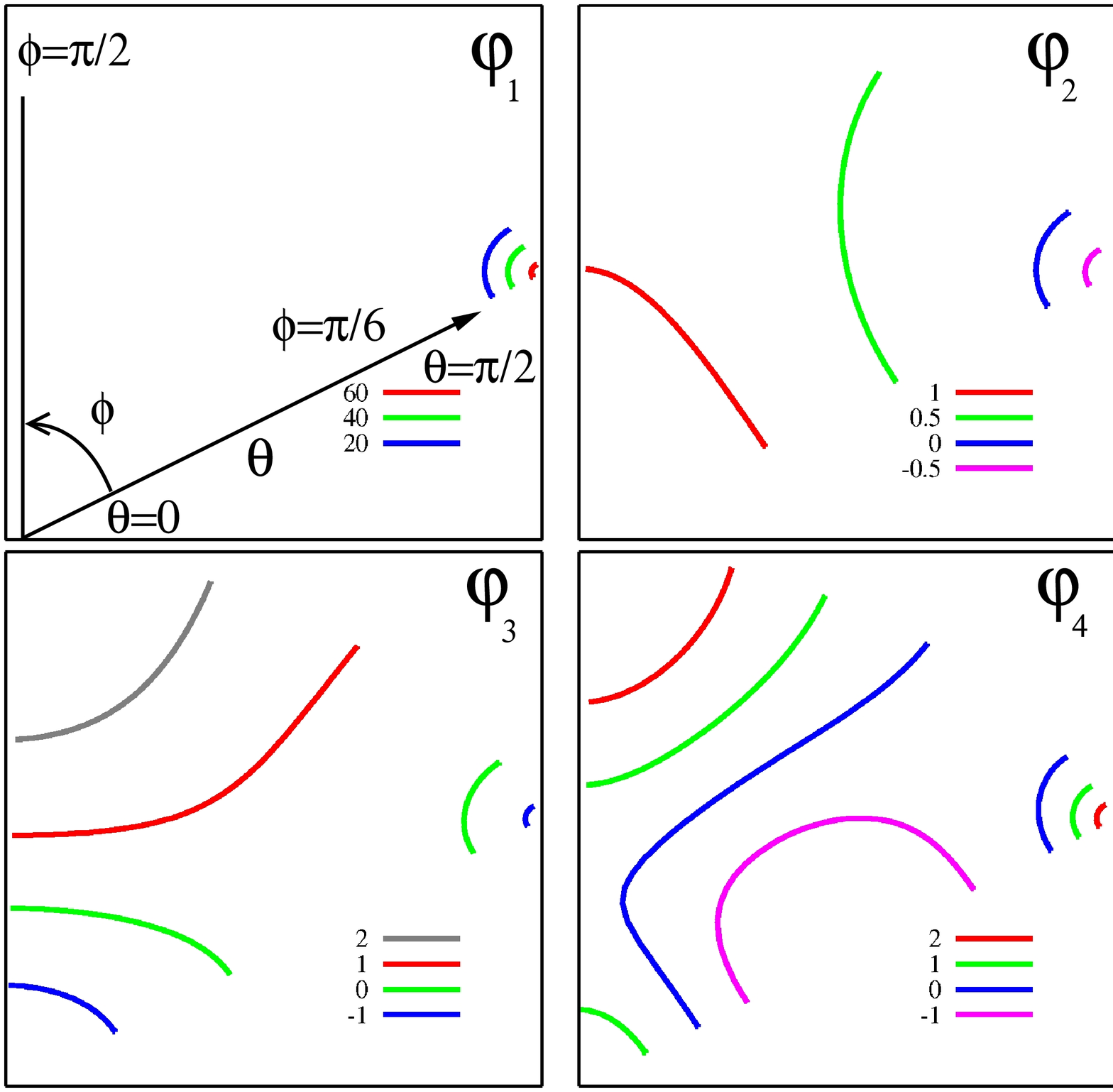}
\caption{(Color online) Contour plots of hyperangular wave functions at $\rho$ = 0.001 a.u. for system with potential given by Eq. (\ref{eq:2body_barrier}). Wave functions are represented in the same way as in Fig. \ref{fig:hpawfs0001}. This value of $\rho$ corresponds to the dissociation region. Note that the wave function of the lowest adiabatic state (left upper panel) is nonzero only in the small region of hyperangular space that represents the dimer plus free atom configuration for large hyper-radius.}
\label{fig:hpawfs001}
\end{figure}

To provide an insight into the hyperangular part of the total wave function, in Figs. \ref{fig:hpawfs0001} and \ref{fig:hpawfs001} we give hyperangular wave functions calculated for two different values of hyper-radius, $\rho= 0.001$ and 0.01 a.u., correspondingly. For the small hyper-radius, the adiabatic states are delocalized. For the large hyper-radius, the lowest state is strongly localized: the amplitude is not zero only in the region where $\phi\sim\pi/6$ and $\theta\sim\pi/2$, corresponding to a dimer + free particle dissociation. Other adiabatic states at the large hyper-radius correspond to the three-body break-up: the corresponding wave functions are less delocalized, that means different possible rearrangements of the final dissociation products.

\section{Example: Efimov resonances in a model three-body system}
\label{sec:Efimov}

The second system we consider is three interacting identical bosons that can form long-range quasi-bound states \cite{nielsen2002}, so-called Efimov resonances. The bosons have masses $m=1$ a.u. and interact through the two-body short-range potential
\begin{equation}
V_2(r)=-17.796 \exp(-r^2).
\end{equation}
Two bosons with the interaction potential have a deeply bound state with energy $E^{(2)}_1=-7.153$ a.u. and a weakly bound state with energy  $E^{(2)}_2=-1.827\times 10^{-9}$ a.u. \cite{nielsen2002}. The $s-$wave scattering length is 23,394.87 a.u.  \cite{nielsen2002}. The system of three bosons interacting through the potential with such a large scattering length and a relatively small radius of forces $r_0\sim 1$ is known to have a number of weakly bound three-body states \cite{efimov1971}. Because for this particular potential there is the deeply bound two-body state, then weakly bound three-body states can decay into a dimer and a free particle. Therefore, the three-body states are indeed resonant states with finite lifetimes (and widths).

Nielsen {\it et al.}  \cite{nielsen2002} determined positions and widths of these states using an $R-$matrix type approach and the hyper-radial coordinates.  Instead of solving the eigenvalue problem along the hyper-radius, in Ref. \cite{nielsen2002} the three-body scattering problem was considered. Namely, the energy-dependent scattering matrix $S(E)$ was obtained for energies above the lowest dissociation limit (deeply bound dimer + a free particle). For the present situation when there is only one open channel, $S(E)$ is a scalar function of $E$: $S=e^{2i\delta(E)}$. The positions and widths of the resonances were obtained considering the phase-shift $\delta(E)$. We use the results by Nielsen {\it et al.}  \cite{nielsen2002}  to test the present method.

Since the Efimov states are very weakly bound, their wave functions extend to very large distances. For example, the minimum of the lowest adiabatic state is around 1 a.u., but the exponentially decaying tail in the closed channel of the fourth Efimov resonance reaches the distance of $10^5$ a.u. (see Fig. 1 of Ref. \cite{nielsen2002}). At the same time, the component of the wave function corresponding to the open channel oscillates with the wavelength of $\lambda=2.5$ a.u. when hyper-radius is large. Therefore, in order to represent the complete wave function, one has to have a grid  with a large number of points. Nielsen {\it et al.}  \cite{nielsen2002} used about 3 million grid points in the interval $0.1<\rho<10^6$ a.u. 

Our method allows us to use a much smaller grid. The exponentially decaying tail can be represented with a modest number of grid points (DVR points in our method). This can be made with a variable grid step, which grows logarithmically with hyper-radius. However, to be able to use the logarithmically-growing step, one has to absorb the outgoing dissociative flux at a small distance where the grid step is still small and the oscillating component of the wave function is still represented properly. Therefore, we place the absorbing potential at the lowest adiabatic potential only at a small distance. 

\begin{table}[tbp]
\vspace{0.3cm} 
\begin{tabular}{|p{2. cm}||p{3.5cm}|p{3.5cm}|p{3.5cm}|p{3.cm}|}
\hline
 & $E_1$, units of $10^{-2}$ &  $E_2$, units of $10^{-4}$  &  $E_3$, units of $10^{-7}$ &  $E_4$, units of $10^{-9}$\\ 
\hline
Ref.  \cite{nielsen2002}    & $-94.14-i2.4705$     &  $-28.38-i3.001$     &$-58.42-i5.79$   & $-19.92-i1.771$ \\ 
\hline
4 ad. states                &    $-96.0- i 2.38$   &    $ -28.8  - i 2.91$  & $ -56.7 -i 5.59$ & $-22.6-i 1.67$\\ 
\hline
5 ad. states                &    $-96.0- i 2.39$   &    $ -29.1  - i 2.98$ & $ -57.2 -i 5.72$& $-22.5-i 1.76$\\ 
\hline
6 ad. states                &    $-96.0- i 2.39$   &    $ -29.2  - i 3.00$ & $ -56.3 -i 5.75$& $-22.6-i 1.77$\\ 
\hline
\end{tabular}
\vspace{0.3cm}
\caption{Comparison of complex energies (in atomic units) obtained in this work using different number of included adiabatic states with the results of Nielsen {\it et al.} \cite{nielsen2002}. The imaginary part of the energies is the halfwidth of the resonances. The overall agreement is good except for the position of the 4$^{th}$ resonance, which is probably not represented accurately in this study: the grid is not long enough in the present calculation. In Ref. \cite{nielsen2002}, six adiabatic states have been used.}
\label{table:gauss_energies}
\end{table}

\begin{figure}[ht]
\includegraphics[width=10cm]{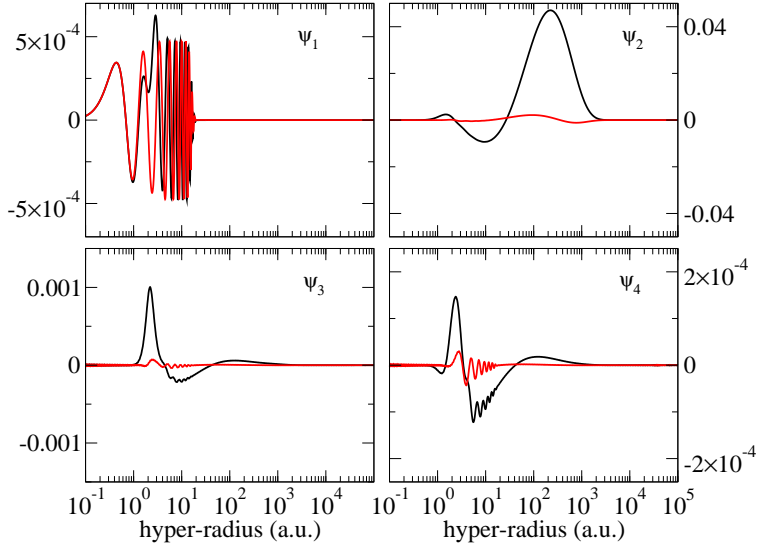}
\caption{(Color online) First four components $\psi_a(\rho)\, (a=1-4)$ of the hyper-radial wave function of the third Efimov resonance. Real and complex parts of the components are shown in black and red lines. The main contribution to the wave function is from the second adiabatic $\psi_a$ state having a minimum around 2 a.u. The only adiabatic state open for dissociation is the ground state, $a=1$. However, the components with $a=2-4$ have small oscillating tails (may not be visible), which are due to the coupling of the corresponding adiabatic states to $\varphi_1(\rho; \theta,\phi)$. This is a generic property of the hyperspherical adiabatic approach: the couplings between adiabatic states decays slowly with $\rho$. Beyond $\rho=20$ a.u., the oscillations on the lowest adiabatic state are damped due to the presence of the absorbing potential.}
\label{fig:wf_gauss_res3}
\end{figure}

In a typical calculation, the variable grid in $\rho$ has only 340 points and extends from 0.002 to 20,000 a.u. The grid step is constant from  $\rho=0.002$ to 0.1 a.u. This region corresponds to the minimum of the two lowest adiabatic states (Fig. 1 of Ref. \cite{nielsen2002}). From $\rho=1$ a.u. the grid step changes logarithmically with $\rho$: $\Delta\rho = 0.04\rho$.  We use a quadratic absorbing potential placed at the lowest adiabatic state. CAP starts at $\rho=$12 a.u; its length is $L=10$ a.u., and the strength $A_2=11$ a.u. The parameters $L$ and $A_2$ are chosen  according to Ref. \cite{vibok1992} in order to minimize reflection and transmission coefficients for the outgoing dissociation flux. In the calculation we used only 4 adiabatic states. The calculation of the four adiabatic states and the corresponding wave functions at 340 grid points of $\rho$ is made in parallel on 96 processors and took about an hour. The size of the matrix for the hyper-radial eigenvalue problem  is $(340*4)\times(340*4)$. The construction and diagonalization of the matrix is made on a single processor and takes also about one hour. With the modest calculation effort, we obtained all four Efimov resonances with a reasonable accuracy. Table \ref{table:gauss_energies} compares calculated energies and widths of the Efimov resonances with the results from Ref. \cite{nielsen2002}. To give an idea about the wave function of an Efimov resonance, Fig. \ref{fig:wf_gauss_res3} shows the four components of the third resonance. The wave function 'lives' mainly in the first excited adiabatic state: the probability to find the system in the first excited state is the largest. Regular oscillations are present mainly in the lowest adiabatic state, which corresponds to the open dissociation path. Beyond 20 a.u., the oscillations decay because of the complex absorbing potential. Smaller oscillations visible in other adiabatic states are due to the coupling to the lowest state. Oscillations along the states decay with the distance $\rho$ because the coupling becomes smaller and the system can dissociate only through the lowest state. 

\section{Summary and conclusions}
\label{sec:concl}

We developed a method to calculate three-body bound states and resonances. The method uses the adiabatic hyperspherical approach, slow variable discretization, sine DVR basis set, and a complex absorbing potential placed  at a large hyper-radius. Using the hyper-radius as a dissociation coordinate allows us to account for two-body and three-body breakup uniformly. If needed, the three-body and two-body dissociation channels could be distinguished by inspecting the asymptotic behavior of resonance wave functions. To simplify the calculation, we employed a variable grid step in the two hyperangles and hyper-radius. In practice, the optimal variable change should be adapted to the given three-body potential.

We applied the developed method to two different three-body systems of identical particles interacting through pairwise potential. However, the method works also for distinguishable particles as well as for intrinsically three-body potentials. The first system consisted of particles interacting through the two-body potential with a barrier. Our calculation for the only bound state is in agreement with the previous study \cite{fedorov03}, where the Faddeev equations were used. However, the position and width of the only resonance does not agree with Ref. \cite{fedorov03}. The reason for this disagreement is not clear:  we believe that our result is correct due to the larger size of the grid that we used. The second system considered here has four weakly bound resonances: Efimov resonances \cite{nielsen2002}. The widths and positions of the resonances obtained in the present study are in good agreement with results of Ref. \cite{nielsen2002}.

Among the advantages of the developed method is a relatively modest basis set along the hyper-radius. For example, in their $R$-matrix method Nielsen {\it et al.} had to use several million grid points along the hyper-radius to obtain converged results. We needed only 340 points. For a better precision, this number could be doubled. Another advantage is that the method does not rely on the evaluation of non-adiabatic coupling terms, which are difficult to represent accurately on a  coarse hyper-radial grid. Most of the calculation can be done on a parallel computer, only the last step, which is relatively short, should be done on a single processor. This is an important advantage in view of the availability of parallel computers. The method works for intrinsically three-body potential and can be easily adapted for distinguishable particles.

The method can be used in a number of problems dealing with long-living states of systems with several degrees of freedom, if a dissociation coordinate could be separated from other coordinates. Such a dissociation coordinate should be used as an adiabatic coordinate. The other coordinates could be treated in the same way as the two hyperangles in the present approach. Depending on how good is the 'adiabaticity' of the dissociation coordinate, the number of adiabatic channels to include could be different: for the dissociation coordinate that is not adiabatic (slow), one may need to include many channels for convergence. Among possible applications of the method could be two-electron or electron-positron problems, pre-dissociation of small polyatomic molecules, decay of nuclei, processes in ultra-cold atomic gases. For example, applications to realistic atomic systems such as Efimov resonances in cesium \cite{kraemer2006} will allow the exploration of the possibility of stabilizing the Efimov resonances into deeply bound levels by Raman processes. Such processes would allow the formation of stable triatomic molecules. The present method provides accurate three-body wave functions, which are essential to compute the radiative transition probability.

\begin{acknowledgments}

Acknowledgment is made to the Donors of the American Chemical Society Petroleum Research Fund for support of this research. This work has been partially supported by the National Science Foundation under Grant No. PHY-0427460, by an allocation of NCSA and NERSC supercomputing resources (project \# PHY-040022). We would also like to thank the Florida Education Fund, which helped make this work possible through the McKnight Doctoral Fellowship. V.K. also acknowledges the support and hospitality of Laboratoire Aim\'e Cotton, Orsay and Fritz Haber Research Institute, Jerusalem. This work was initiated when V.K. was visiting these laboratories.

\end{acknowledgments}


\begin{thebibliography}{99}

\bibitem{efimov1971} V.~Efimov, Sov. J. Nucl. Phys. {\bf 12}, 589  (1971); V.~Efimov, Sov. J. Nucl. Phys. {\bf 29}, 546  (1979).

\bibitem{esry96} B.~D.~Esry, C.~D.~Lin, and C.~H.~Greene, Phys. Rev. A {\bf 54}, 394 (1996).

\bibitem{braaten2002} E.~Braaten, H.-W.~Hammer, and M.~Kusunoki, Phys. Rev. Lett. {\bf 90}, 170402 (2003); E.~Braaten, H.-W.~Hammer Phys. Rep. {\bf 428}, 259 (2006). 

\bibitem{stoll2005} M.~Stoll, T.~Kohler, Phys. Rev. A {\bf 72}, 022714 (2005).

\bibitem{esry2006}  L.~P.~D'Incao and B.~D.~Esry, Phys. Rev. Lett. {\bf 94}, 213201 (2005); L.~P.~D'Incao and B.~D.~Esry, Phys. Rev. A {\bf 72}, 032710 (2005); L.~P.~D'Incao and B.~D.~Esry, Phys. Rev. A {\bf 73},  030702  (2006);  L.~P.~D'Incao and B.~D.~Esry, Phys. Rev. A {\bf 73}, 030703 (2006); B.~D.~Esry and C.~H.~Greene, Nature {\bf 440}, 289 (2006).

\bibitem{kraemer2006} T.~Kraemer, M.~Mark, P.~Waldburger, J.~G.~Danzl, C.~Chin, B.~Engeser, A.~D.~Lange, K.~Pilch, A.~Jaakkola, H.-C.~N\"agerl, and R.~Grimm, Nature {\bf 440}, 315 (2006).

\bibitem{penkov99} F.~M.~Pen'kov, Phys. Rev. A {\bf 60}, 3756 (1999).

\bibitem{nielsen2002} E.~Nielsen, H.~Suno, and B.~D.~Esry, Phys. Rev. A {\bf 66}, 012705  (2002).

\bibitem{tolstikhin96} O.~I.~Tolstikhin, S.~Watanabe, and M.~Matsuzawa, J. Phys. B: At. Mol. Opt. Phys.  {\bf 29}, L389 (1996).

\bibitem{kokoouline2006} V.~Kokoouline and F.~Masnou-Seeuws, Phys. Rev. A {\bf 73}, 012702  (2006).

\bibitem{borisov01} A.~G.~Borisov, J. Chem. Phys., {\bf 114}, 7770  (2001).

\bibitem{willner2004} K.~Willner, O.~Dulieu, and F.~Masnou-Seeuws, J. Chem. Phys. {\bf 120}, 548  (2004); K.~Willner, Ph.D. thesis, Universit\'e Paris Sud, 2005.

\bibitem{fedorov03} D.~V.~Fedorov, E.~Garrido, and A.~S.~Jensen, Few-Body Sys. {\bf 33}, 153 (2003).

\bibitem{johnson80}  B.~R.~Johnson, J. Chem. Phys. \textbf{73}, 5051 (1980).

\bibitem{kokoouline03} V.~Kokoouline and C.~H.~Greene, Phys. Rev. A {\bf 68}, 012703 (2003).

\bibitem{kokoouline99} V.~Kokoouline, O.~Dulieu, R.~Kosloff, and F.~Masnou-Seeuws, J. Chem. Phys. {\bf 110}, 9865  (1999).

\bibitem{vibok1992} \' A.~Vib\' ok and G.~G.~Balint-Kurti, J. Phys. Chem. {\bf 96},\hspace{0.25em}8712  (1992).

\bibitem{riss1993} U.~V.~Riss and H.-D.~Meyer, J. Phys. B: At. Mol. Opt. Phys. {\bf 26}, 4503  (1993).

\bibitem{tiemann86} E.~Tiemann, Molec. Phys. {\bf 65}, 359 (1988).

\bibitem{davydov} A.~S.~Davydov, {\it Quantum Mechanics} (Pergamon, Oxford, 1965), p. 472.


\end{thebibliography}

\end{document}